\documentclass[twocolumn]{autart}    

\usepackage{amsmath, bm}
\usepackage{amssymb}
\usepackage{commath}
\usepackage{graphicx}          

\newcounter{as}
~\itshape\ignorespaces
{\par\ignorespacesafterend}

\usepackage{enumitem}
\newcommand{\subscript}[2]{$#1 _ #2$}

\usepackage{color,xifthen}
\newcommand{\displaycomments}
\makeatletter
\ifdefined\displaycomments
\newcommand{\FIXME}[1][]{%
\ifthenelse{\isempty{#1}}{%
\noindent\textcolor{red}{\emph{FIXME}}%
}{%
\noindent\textcolor{red}{\emph{FIXME: {#1}}}%
}%
}%
\else
\newcommand{\FIXME}{}%
\fi
\makeatother

\begin{document}

%

\begin{frontmatter}

\title{Experiment Design for Identification of Marine Models\thanksref{footnoteinfo}} 

\thanks[footnoteinfo]{Corresponding author F.~Ljungberg.}

\author[liu]{Fredrik Ljungberg}\ead{fredrik.ljungberg@liu.se},    			
\author[abb]{Jonas Linder}\ead{jonas.x.linder@se.abb.com},           		
\author[liu]{Martin Enqvist}\ead{martin.enqvist@liu.se}\ and \ 				
\author[abbmarine]{Kalevi Tervo}\ead{kalevi.tervo@fi.abb.com}  				

\address[liu]{Linköping University, Linköping, Sweden}  								
\address[abb]{ABB Corporate Research, Västerås, Sweden}             					
\address[abbmarine]{ABB OY, Industrial Automation, Marine and Ports, Helsinki, Finland} 


\begin{keyword}                             
System identification; experiment design; nonlinear models; greybox modelling; marine systems.                
\end{keyword}

\begin{abstract}                          
In this work, experiment design for marine vessels is explored. A dictionary-based approach is used, $i.e.$, a systematic way of choosing the most informative combination of independent experiments out of a predefined set of candidates. This idea is quite general but is here tailored to an instrumental variable (IV) estimator with zero-mean instruments. This type of estimator is well-suited to deal with parameter estimation for second-order modulus models, which is a class of models often used to describe motion of marine vessels. The method is evaluated using both simulated and real data, the latter from a small model ship as well as from a full-scale vessel. Further, a standard motion-planning problem is modified to account for the prior-made choice of information-optimal sub-experiments, which makes it possible to obtain a plan for the complete experiment in the form of a feasible trajectory.
\end{abstract}

\end{frontmatter}

\section{Introduction}
Ship motion and control have engaged researchers for at least a century, see for example \cite{minorsky1922directional} for an early reference. Even if the controller concept has hardly changed since then, there have been great advances made. One major difference in modern-time automatic steering of ships is that many control methods are model based. Therefore, having accurate simulation models available is becoming increasingly important as ships are becoming more automated. There are essentially two classes of experiments that can be employed to generate data that such models can be based on. Firstly, there are experiments where motion-yielding forces are produced by some external equipment, such as towing-tank tests. Very accurate parameter estimates can be obtained in this way, but the experiments are usually expensive and time-consuming. Moreover, since a down-scaled replica is often used in this type of experiment, modelling errors due to mismatches between the scale model and the full-size vessel are common \cite{yoon2003identification}. An often cheaper way of obtaining models is through system identification, $i.e.$, by modelling based on data collected from tests where the vessel's own propulsion system is used to generate the motion. 

It is well known that the choice of input signal during the data acquisition is a significant factor for the result of the parameter estimation in system identification, see for example \cite{ljung1999system}. For a data-collection experiment, there are both practical and statistical considerations to be made. The practical considerations relate to physical limitations, for example, traffic regulations, actuator saturations and avoiding of obstacles such as shallow water, islands and other vessels. The statistical aspect of the design relates to maximizing the information gained from the experiment and can often be studied in a more general and abstract fashion. Input design for general linear time-invariant systems has been studied for some time and is by now a fairly well-understood topic, see for example \cite{goodwin1977dynamic} for an early reference and \cite{bombois2011optimal} for a more recent overview. However, accurate ship modelling often requires that nonlinear hydrodynamic forces and moments are taken into account if the model should cover the ship's full speed envelope. Experiment design for nonlinear systems is not yet as well understood, and most previous works have dealt with experiment design for certain subclasses of nonlinear systems. In an early work regarding experiment design for nonlinear systems, \cite{hjalmarsson2007optimal} discussed a two-step approach to solve the input-design problem, one step in which the optimal probability density function of the input is computed and one in which the signal is realized based on the found probability density. This idea is inspired by how experiment design is commonly performed for linear systems and was extended in \cite{larsson2010optimal}. There, input design for general nonlinear finite-input-response (NFIR) models was considered, a model class which was also dealt with in \cite{decock2016}. Another often studied subclass of nonlinear systems is interconnections of linear systems and static nonlinearities. Experiment design for such systems was explored in \cite{mahata2016information,vincent2010input}. Further, for static nonlinear systems, which are linear in the parameters, the input design can be formulated as a convex optimization problem. This property was used in \cite{forgione2014experiment} to perform input design for dynamical systems by considering a fixed number of amplitude levels of the input signal. A similar approach was taken in \cite{valenzuela2015graph}, where input design for nonlinear output-error (NOE) models was considered using graph theory. In \cite{gopaluni2011input}, a particle-filter approach was used and in \cite{umenberger2019nonlinear}, the input-design problem was transformed to an optimal control problem using tools from statistical inference. Both these approaches consider fairly general classes of nonlinear models but rely on stochastic approximation techniques and require initial guesses for the input signals to deal with the resulting non-convex formulations. 

For most model structures, the optimal experiment design will depend on the parameters of the system to be identified, $i.e.$, the quantities that it is desired to estimate. This can seem a bit restrictive, but for many applications there is some nominal model available that the experiment design can be based on. In other cases a multi-stage approach can be employed, where an estimation step and a design step are alternated iteratively, as in \cite{gerencser2009identification}. A third option is a robust design, where the experiment-design criterion is optimized over a distribution of possible parameter values. This last approach was explored for nonlinear models in \cite{valenzuela2017robust}.

Parameter estimation for nonlinear models of industrial manipulators was discussed in~\cite{wernholt2011nonlinear}. A two-step solution was used, where frequency-response estimates for local linear models were first estimated for a chosen number of operating points. Suitable values for parameters of a nonlinear model were then found in a second step by minimizing the discrepancy between these locally valid frequency-response estimates and the parametric frequency response of the nonlinear model. Experiment design for such an approach was explored in \cite{wernholt2007experiment}. In this case, the design reduces to a selection between a predefined set of operating points, where tools from experiment design for linear systems can be applied for finding a suitable input signal for each operating point. This is appealing because the Fisher information matrix of a sequence of sub-experiments can be written as a convex combination of the Fisher information matrices of the individual sub-experiments. As a result, the optimal experiment-design problem is convex and can be solved efficiently.

The importance of collecting informative data has also been stressed in works regarding marine modelling. In \cite{blanke1999optimized,blanke2006efficient}, surface vessels were considered and the informativity gains from zig-zag, circle and spiral maneuvers as well as telegraph input signals, were compared. Further, an optimal zig-zag maneuver was found in \cite{yoon2003identification} by empirically trying out different candidates of varying amplitude and frequency. More recently, an approach for finding optimized input signals for nonlinear models of underwater vehicles was presented in \cite{nouri2018optimal}. Their proposed idea was to parameterize a step-wise changing signal in terms of a finite number of amplitude levels and dwell times, $i.e.$, an approach similar to \cite{forgione2014experiment,valenzuela2015graph}. A comparable approach was applied for surface vessels in \cite{wang2020optimal}. Both \cite{nouri2018optimal} and \cite{wang2020optimal} obtain optimization problems that can be solved in reasonable time by imposing restrictions on the input signal, such as having the propeller thrust fixed and only varying the rudder angle. In both cases, the resulting optimization problems are solved using numerical methods and the optimized input signals are in simulation experiments shown to give more accurate parameter estimates than zig-zag maneuvers. Lastly, the benefits of having the vessel's own propulsion system generate motion during data collection were shown in a comparative study between model estimation following towing tests and self-actuated tests for an underwater vehicle in~\cite{lack2019experimental}. 

As seen in the references above, data for estimation of marine models are often collected using standard maneuvers, such as turning circles, spiral motions and zig-zag tests. These maneuvers were not initially developed for generating informative data but for evaluating a ship's maneuvering capabilities~\cite{wang2020optimal}. A standard maneuver alone does not necessarily excite all relevant system modes, but when multiple maneuvers are performed in sequence, informative data can often be obtained. Based on this observation and inspired by \cite{wernholt2007experiment}, this work evaluates a dictionary-based approach for experiment design of marine vessels, $i.e.$, a systematic way of choosing the most informative combination of independent sub-experiments out of a predefined set of candidates. This idea is tailored to an instrumental variable (IV) estimator, which due to high accuracy under general disturbance conditions has been shown to work well for models of marine vessels in \cite{ljungberg2022a}. There, it was noted that choosing instruments with zero mean can be of importance for an IV estimator when applied to such models. Consequently, a technique to account for this in the experiment design is suggested. The proposed method is evaluated in an experimental study using both simulated and real data from a scale ship as well as a full-size marine vessel. Lastly, modifications are made to a standard motion-planning problem to account for the prior-made choice of information-optimal sub-experiments. This makes it possible to obtain a plan for the complete experiment in the form of a feasible trajectory and the potential of the idea is shown in a simulation example, where the dictionary-based method for choosing a set of sub-experiments is supplemented with a lattice-based motion planner.

One contribution of this work is the adaptation of the dictionary-based approach in \cite{wernholt2007experiment} to time-domain estimation using an IV~estimator with zero-mean instruments. Another contribution is the observation that there are multiple conceivable ways of subtracting the instrument mean when data is collected from different sub-experiments and that this choice is important for the experiment design. Additionally, the modifications to the motion-planning problem for (indirectly) taking informativity into account are novel. 

In the following, some brief theoretical preliminaries are given in Section~\ref{sec:prob_form}. After that, the considered dictionary-based approach for experiment design is outlined in Section~\ref{sec:basic_method}. This approach is then adapted to work well for marine models in Section~\ref{sec:marine_method}, where the results from experimental work are also shown. Further, the method for generating a plan for the complete experiment in the form of a feasible trajectory is presented in Section~\ref{sec:gen_traj}. Lastly, the paper is concluded in Section~\ref{sec:conc}.

\section{Problem formulation and preliminaries} \label{sec:prob_form}
Accurate ship modelling requires that nonlinear hydrodynamic forces and moments are taken into account if the model should cover the ship's full speed envelope. To this end, consider a nonlinear discrete-time state-space system
\begin{subequations}
	\begin{align} 
		\bm{x}(k+1) &= \bm{f}(\bm{x}(k), \bm{u}(k), \bm{\theta}_0) + \bm{w}(k), \\
		\bm{y}(k) &=  \bm{x}(k) + \bm{e}(k),
	\end{align}
\end{subequations}
where $\bm{u}(k) \in \mathbb{R}^{n_u}$ is a known input signal, $\bm{x}(k) \in \mathbb{R}^{n_x}$ is a vector consisting of the latent system states, all of which are measured directly (with noise) and collected in the output vector $\bm{y}(k)  \in \mathbb{R}^{n_x}$. Moreover, $\bm{e}(k)  \in \mathbb{R}^{n_x}$ and $\bm{w}(k)  \in \mathbb{R}^{n_x}$ are disturbance signals and $\bm{\theta}_0  \in \mathbb{R}^{n_{\theta}}$ is a vector of parameters, which is assumed to not vary over time. 

\subsection{Instrumental variable estimator}
For a predictor model on regression form
\begin{align} 
	\hat{\bm{y}}\left(k, \bm{\theta}\right) = \bm{\Phi}^T(k) \bm{\theta},
\end{align}
the IV estimate is defined as the solution to the optimization problem
\begin{align} \label{eq:IV_estimator_def}
	\hat{\bm{\theta}}_N^{IV} = \underset{\bm{\theta}}{\text{argmin}} \ V_N^{IV}(\bm{\theta}),
\end{align}
where the cost function is
\begin{align} \label{eq:IV_estimator_def_obj_fun}
	V_N^{IV}(\bm{\theta}) = \left\Vert \frac{1}{N} \sum_{k=1}^N   \bm{Z}(k) \left(\bm{y}(k)-\bm{\Phi}^T(k) \bm{\theta} \right) \right\Vert_{2}^2.
\end{align}
Here $\bm{Z}(k) \in \mathbb{R}^{n_{\theta} \times n_x}$ is called the instrument matrix and plays an important role for the accuracy of the resulting models. Ideally, the instrument matrix should be correlated with the system states but uncorrelated with the disturbances. The consistency of an IV estimator was shown in \cite{ljungberg2022a} for second-order modulus models, $i.e.$, models with signed-square regressors, such as $x\abs{x}$. This model class is often used for describing motion of marine vessels. Consequently, the IV estimator will be employed in this work as well. 

\section{Basic method} \label{sec:basic_method}
If the system can be fully described by the model structure, it is possible to evaluate the usefulness of an estimator based on the bias and covariance of the parameter estimates. In this paper, the focus is on the covariance since the consistency of the estimator was explored in \cite{ljungberg2022a}. A common starting point when formulating an experiment-design problem for minimizing the variance is to assume that the estimator is asymptotically efficient, which means that the covariance matrix of the estimated parameters will converge to the inverse of the Fisher information matrix as the amount of estimation data increases, see for example \cite{ljung1999system}. In this case, making the covariance matrix as small as possible is equivalent to making the information matrix as large as possible, at least asymptotically. This is appealing because it is usually easier to obtain an expression for the information matrix, which to a large extent depends on the Hessian of the cost function, in comparison to directly finding an analytical expression for the covariance matrix. Despite the fact that the IV method gives increased possibilities of obtaining consistency, IV estimators are in general not asymptotically efficient. Therefore, the conventional arguments in terms of Fisher information do not apply to the estimators studied in this work. However, maximizing the Hessian of the cost function with respect to the parameters is useful from a practical point of view since it makes the output of the estimator sensitive to changes in the parameters. Thereby, performing experiments in such a way should in general give more distinct optima to the resulting parameter estimation problems. To this end, let
\begin{align}
	\bm{G}(N) &\overset{\Delta}{=} \frac{1}{2} \frac{\partial^2}{\partial \bm{\theta}^2} V_N^{IV}(\bm{\theta}),
\end{align}
denote the Hessian of the cost function and note that
\begin{align*}
	\bm{G}(N) &= \left[\frac{1}{N} \sum_{k=1}^N \bm{\Phi}(k)  \bm{Z}^T(k)\right] \left[\frac{1}{N} \sum_{k=1}^N \bm{\Phi}(k)  \bm{Z}^T(k)\right]^T.
\end{align*}
Even if this is technically not the Fisher information matrix it will subsequently be referred to as the information matrix. The fact that maximizing $\bm{G}(N)$ reduces the uncertainty of the parameter estimates is shown with a simulation example in Section~\ref{sec:simulation_study}.

Ideally, the optimization problem
\begin{equation*}
	\begin{aligned}
		\bm{u}^* = & \underset{ \bm{u}(1), \hdots, \bm{u}(N)}{\text{argmax}}
		& &  \ell \left(  \bm{G}(N) \right)\\
		& \text{ s.t.}
		& & \bm{u}(1), \hdots, \bm{u}(N) \in \ \text{Feasible input}, \\
		& & & \bm{y}(1), \hdots, \bm{y}(N) \in \ \text{System dynamics},
	\end{aligned}
\end{equation*}
would be solved for some scalar function $\ell(\cdot)$. However, this is in general a non-convex problem and solving it would require a good initial guess for the input signal. Such an initial guess is not that easy to obtain for marine vessels because they are often equipped with multiple thrusters, which means that there are many degrees of freedom (DOF) in the input design. A simpler alternative is to use a dictionary-based approach where a sub-optimal input sequence is constructed by selecting the most informative combination of input signals out of a predefined set. An important step of such an approach is the selection of signals in the dictionary because if the dictionary does not include candidate signals that can be combined to obtain informative data, the optimized design will not be informative either. 

Here, it is assumed that candidate signals with this property exist and that they are represented by the set
\begin{align}
	\mathcal{U} = \left\{\bm{u}_1(k), \hdots,  \bm{u}_Q(k)\right\}.
\end{align}
Subsequently, the motion trajectories obtained when applying these will be referred to as experiment primitives. Moreover, introduce the variable
\begin{align}
	\tilde{\bm{\Phi}}(k) = \bm{\Phi}(k)  \bm{Z}^T(k),
\end{align}
for simplified notation and let
\begin{align}
	\bm{G}_q(N_q) = \left[\frac{1}{N_q} \sum_{k=1}^N \tilde{\bm{\Phi}}_q(k) \right] \left[\frac{1}{N_q} \sum_{k=1}^N \tilde{\bm{\Phi}}^T_q(k) \right],
\end{align}
be the Hessian of the cost function when using the candidate signal $\bm{u}_q(k)$ and instrument matrix $\bm{Z}_q(k)$. If all the candidate input signals are applied in sequence, the total information gain is
\begin{align}
	\bm{G}(N) &=  \bm{\Gamma}(N) \bm{\Gamma}^T(N),
\end{align}
where
\begin{align}
	\bm{\Gamma}(N) &= \frac{1}{N} \left(  \sum_{k=1}^{N_1}  \tilde{\bm{\Phi}}_1(k) + \hdots + \sum_{k=1}^{N_Q}  \tilde{\bm{\Phi}}_Q(k) \right),
\end{align}
and where $N = N_1+\hdots+ N_Q$, $i.e.$, the total experiment time. Since the Hessian is a matrix it is necessary to choose some scalar measure of it to optimize. There are multiple different such criteria that can be used, some of which are discussed in~\cite{mehra1974optimal}. One choice is to maximize the determinant. This is called a D-optimal design and will be used in this work. Geometrically, a D-optimal design is comparable to minimizing the uncertainty volume of the estimated parameters.
Further, the matrix $\bm{\Gamma}(N)$ defined above is symmetric, and consequently, it holds that
\begin{align}
	\det \left(\bm{G}(N) \right) = \det \left(\bm{\Gamma}(N) \bm{\Gamma}^T(N) \right) = \det \left( \bm{\Gamma}(N)  \right) ^2.
\end{align}
Therefore, maximizing the determinant of $\bm{G}(N)$ is equivalent to maximizing the determinant of $\bm{\Gamma}(N)$ if $\bm{\Gamma}(N) \succeq 0$. Since $\bm{\Gamma}(N) \succeq 0$ can not be guaranteed in a general case it is possible to instead maximize $\abs{\det\left( \bm{\Gamma}(N)\right)}$. Now, let 
\begin{align}
	\bm{\Gamma}_q(N_q) = \frac{1}{N_q} \sum_{k=1}^{N_q}  \tilde{\bm{\Phi}}_q(k),
\end{align}
be the information gain from using $\bm{u}_q(k)$ and $\bm{Z}_q(k)$. If stationary signals are considered, it will asymptotically be the case that
\begin{align}
	\lim_{N_q \to \infty} \frac{1}{N_q} \sum_{k=1}^{N_q}  \tilde{\bm{\Phi}}_q(k) \overset{\Delta}{=} \bar{\bm{\Gamma}}_q,
\end{align}
and consequently it holds that
\begin{align}
	N_q \frac{1}{N_q} \sum_{k=1}^{N_q} \tilde{\bm{\Phi}}_q(k)  \approx N_q \bar{\bm{\Gamma}}_q,
\end{align}
for significantly large $N_q$. Under assumption that each candidate signal is applied for a sufficiently long time, it is therefore reasonable to consider the optimization problem
\begin{equation} \label{eq:opt_prob}
	\begin{aligned}
		(N_1^*, \ \hdots, \ N_Q^*) = & \underset{N_1, \hdots, N_Q}{\text{argmax}}
		& & \log \abs{\det \left( \bm{T} \right)} \\
		& \text{ s.t.}
		& & \bm{T} = N_1 \bar{\bm{\Gamma}}_1  + \hdots + N_Q \bar{\bm{\Gamma}}_Q , \\
		& & & N_1 + \hdots + N_Q = N, \\
		& & & N_q \geq 0, \ q=1, \hdots, Q,
	\end{aligned}
\end{equation}
where $\bar{\bm{\Gamma}}_1, \hdots, \bar{\bm{\Gamma}}_Q$ are estimated beforehand, either by initial experiments with the real platform or in a simulation environment with a nominal model. It can be noted that
\begin{align}
	\bm{T} \approx N \bm{\Gamma}(N),
\end{align}
which, since $N$ is not a free variable during the optimization, means that the solution to the problem above will maximize the determinant of $\bm{G}(N)$. The last constraint in \eqref{eq:opt_prob} is a relaxation of forcing each $N_q$ to be a non-negative integer. This relaxation makes the optimization problem easier to solve and has negligible effects on the outcome for reasonably long experiments.

\section{Method adapted for marine vessels} \label{sec:marine_method}
Most nonlinear model structures used for describing motion of marine vessels at speed stem from one of two basic ideas. The first was suggested by~\cite{abkowitz1964lectures} and is based on a Taylor expansion. In this case, the even-order terms are often neglected to enforce that the model behaves in the same way for positive and negative relative velocities, something that is necessary due to ship symmetry. The other type of model structure was first proposed by~\cite{fedyaevsky1964control} and is based on a combination of physical effects such as circulation and cross-flow drag principles. These properties are often well-described by quadratic functions and the constraint of having a symmetric model is therefore instead resolved by use of the modulus function. Models of this type typically do not include any terms of higher order than two and are therefore referred to as second-order modulus models. Moreover, experimental ship data is often collected under presence of environmental disturbances, such as wind and water currents. Correctly dealing with these disturbances is challenging already in the linear case and becomes even harder when models are nonlinear. Doing so is, however, important to not obtain a biased model. 

\subsection{Compensating for zero-mean instruments}
The possibility of obtaining consistency by using an IV estimator for dealing with these challenges was explored in \cite{ljungberg2022a}. In general, the accuracy of an IV estimator is highly dependent on the choice of instruments and a common way of obtaining instruments in practice is by simulation of a nominal model with crude parameter values. In this case, the instrument matrix can be formed as a noise-free version of the regression matrix, see for example \cite{thil2008instrumental}. However, in \cite{ljungberg2022a} it was noted that this was not sufficient and the additional step of forcing the instrument matrix to have zero mean was proposed as a remedy. This turned out to be an efficient way of asymptotically eliminating the influence of certain disturbances on the parameter estimates. When data is collected from different sub-experiments, there are multiple ways in which the instrument mean can be subtracted. One way is to subtract it from one batch of data at a time
\begin{align}  \label{eq:batchwise_subtraction}
	\tilde{\bm{Z}}_q(k) &= \bm{Z}_q(k) - \frac{1}{N_q} \sum_{k=1}^{N_q} \bm{Z}_q(k).
\end{align}
Another alternative is to subtract the mean of the complete dataset at once
\begin{align} \label{eq:complete_subtraction}
	\tilde{\bm{Z}}_q(k) &= \bm{Z}_q(k) - \underbrace{\frac{1}{N} \sum_{q=1}^Q  \sum_{k=1}^{N_q} \bm{Z}_q(k)}_{\overset{\Delta}{=} \bar{\bm{Z}}}.
\end{align}
It turns out that the latter option often gives a reduced variance for the estimated parameters and that better results can be obtained if this is taken into consideration already during the experiment design. The motivation for this is illustrated by the following simple example.  
\begin{exmp}
Consider the data-generating system
\begin{align} \label{the_system}
    y(k) = \theta_0 u(k) + e(k),
\end{align}
and the predictor model
\begin{align}
    \hat{y}(k) =  u(k) \theta \overset{\Delta}{=} \varphi^T(k) \theta.
\end{align}
For a general instrument vector, $\zeta(k)$, the IV estimate is 
\begin{align}
  \nonumber  \hat{\theta}_N^{IV} &= \left(\frac{1}{N} \sum_{k=1}^N \zeta(k) \varphi^T(k) \right)^{-1} \left( \frac{1}{N} \sum_{k=1}^N \zeta(k) y(k) \right) \\
    &= \theta_0 +  \frac{\frac{1}{N}\sum_{k=1}^N \zeta(k)e(k)} {\frac{1}{N} \sum_{k=1}^N \zeta(k) u(k)}.
\end{align}
Further, assume that the input has a time-varying component $\Tilde{u}(k)$ with zero mean and an excitation offset that alters sign halfway through the experiment
\begin{align} \label{the_input}
    u(k) = \begin{cases} \Tilde{u}(k) + \Bar{u}, \ &\text{if} \ k = 1, \hdots N/2, \\
	    \Tilde{u}(k) - \Bar{u}, \ &\text{if} \ k = N/2+1, \hdots N, \end{cases}
\end{align}
Furthermore, assume that $\Tilde{u}(k)$ is repeated in both experiment halves, such that $\Tilde{u}(k + N/2) = \Tilde{u}(k)$ for $k = 1, \hdots N/2$. Now, divide the analysis into two cases. Firstly, consider the case where $\zeta(k) = u(k)$ and note that this corresponds to a complete subtraction of the instrument mean (the mean of $u(k)$ is zero by design). Then, it holds that
\begin{align}
    \hat{\theta}_N^{IV_1} &= \theta_0 +  \frac{\frac{1}{N} \left( \sum_{k=1}^N \Tilde{u}(k) e(k) \right) + \Bar{u} e'_N}{ \frac{1}{N} \left(\sum_{k=1}^N \Tilde{u}(k)^2 \right) + \Bar{u}^2}.
\end{align}
where $ e'_N = \frac{1}{N} \left(\sum_{k=1}^{N/2} e(k) - \sum_{k=N/2+1}^{N} e(k) \right)$.  Secondly, assume that $\zeta(k) = \Tilde{u}(k)$, which corresponds to a batchwise subtraction of the instrument mean. In this case, it holds that
\begin{align}
   \hat{\theta}_N^{IV_2} &= \theta_0 +  \frac{\frac{1}{N} \sum_{k=1}^N  \Tilde{u}(k) e(k)} {\frac{1}{N} \left(\sum_{k=1}^N \Tilde{u}(k)^2 \right)}. 
\end{align}
It can be noted that the variance expression of $\hat{\theta}_N^{IV_1}$ has an additional term, $\bar{u}^2$, in the denominator, which means that it will decrease to zero when $\bar{u}$ increases. Therefore, $\hat{\theta}_N^{IV_1}$ is a better choice in comparison to $\hat{\theta}_N^{IV_2}$ for significant values of $\bar{u}$. 
\end{exmp}

Compensating for a batchwise subtraction of the mean in the experiment design would be straightforward since the relevant signal levels associated with each experiment primitive can be found during the initial experiments where $\bar{\bm{\Gamma}}_1, \hdots, \bar{\bm{\Gamma}}_Q$ are estimated. However, if the mean is subtracted for the complete experiment at once, the usefulness of each sub-experiment can only be assessed in combination with the other sub-experiments. Compensating for a complete subtraction of the mean in the experiment design is therefore a bit more challenging. If the mean is subtracted as in \eqref{eq:complete_subtraction}, it is the case that
\begin{align}
	\bar{\bm{\Gamma}}_q &= \bar{\bm{X}}_q - \bar{\bm{Y}}_q \bar{\bm{Z}}^T,
\end{align}
where 
\begin{align}
\label{eq:xbar} \bar{\bm{X}}_q  &= \lim_{N_q \to \infty} \frac{1}{N_q} \sum_{k=1}^{N_q}  \bm{\Phi}_q(k) \bm{Z}^T_q(k), \\
\label{eq:ybar} \bar{\bm{Y}}_q	&= \lim_{N_q \to \infty} \frac{1}{N_q} \sum_{k=1}^{N_q}  \bm{\Phi}_q(k).
\end{align}
Similar to how $\bar{\bm{\Gamma}}_q$ was estimated for each experiment primitive earlier, $\bar{\bm{X}}_q$, $\bar{\bm{Y}}_q$ and 
\begin{align}
	\label{eq:zbar} \bar{\bm{Z}}_q &= \lim_{N_q \to \infty} \frac{1}{N_q} \sum_{k=1}^{N_q} \bm{Z}_q(k),
\end{align}
can be estimated for each experiment primitive based on initial experiments or a nominal model. Consequently, the  optimization problem
\begin{equation} \label{eq:opt_prob_zero_mean}
	\begin{aligned}
		(N_1^*, \ \hdots, \ N_Q^*)  = & \underset{N_1, \hdots, N_Q}{\text{argmax}}
		& & \log \abs{ \det \left( \bm{T} - \bm{S} \bar{\bm{Z}}^T \right)} \\
		& \text{ s.t.}
		& & \hspace{-8mm} \bm{T} = N_1 \bar{\bm{X}}_1 + \hdots + N_Q \bar{\bm{X}}_Q , \\
		& & & \hspace{-8mm} \bm{S} = N_1 \bar{\bm{Y}}_1  + \hdots + N_Q \bar{\bm{Y}}_Q, \\
		& & & \hspace{-8mm} \bar{\bm{Z}} = \frac{1}{N} (N_1 \bar{\bm{Z}}_1 + \hdots + N_Q \bar{\bm{Z}}_Q), \\
		& & & \hspace{-8mm} N_1 + \hdots + N_Q = N, \\
		& & & \hspace{-8mm} N_q \geq 0, \ q=1, \hdots, Q,
	\end{aligned}
\end{equation}
can be considered as an alternative to \eqref{eq:opt_prob}, where the fact that the first-order moment of the instrument matrix will be removed during the parameter estimation is acknowledged in the experiment design. This optimization problem is non-convex. However, no initial guess for the optimal input signal is required unlike many other non-convex problem formulations used for experiment design. Instead, only a guess for how long the candidate signals should be applied with respect to each other is needed. This is a significantly milder requirement and based on experimental work, it seems to be sufficient to assume that all candidate signals are used equally much.

\subsection{Results from simulation experiments} \label{sec:simulation_study}
In order to illustrate the potential of the proposed method, simulation experiments were performed. These were carried out using a maneuvering model of a marine surface vessel given by
\begin{align*}
	 u(k+1) &= u(k) + \mathcal{X}_u u_r(k) + \mathcal{X}_{u|u|}  u_r(k) \abs{u_r(k)} \\
	 & \hspace{2cm} + \mathcal{X}_{vr} v_r(k)r(k) + \mathcal{X}_{\tau} \tau_1(k), \\
	 v(k+1) &= v(k) + \mathcal{Y}_v v_r(k) + \mathcal{Y}_{ur} u_r(k)r(k) + \mathcal{Y}_{\tau} \tau_2(k), \\
	 r(k+1) &= r(k) + \mathcal{N}_r r(k) + \mathcal{N}_{uv} u_r(k)v_r(k) +  \mathcal{N}_{\tau} \tau_3(k), \\
	\bm{y}(k) &= \begin{bmatrix} u(k) & v(k) & r(k) \end{bmatrix}^T + \bm{e}(k).
\end{align*}
Here, $u(k)$, $v(k)$ and $r(k)$ constitute surge, sway and yaw rate, respectively, and $\bm{\tau}(k) = [\tau_1(k), \ \tau_2(k), \ \tau_3(k)]^T$ are input forces and moments caused by the ship's actuators. The subscripted $r$ signifies relative velocity such that $u_r(k) =u(k) -u_c(k)$ and $v_r(k) = v(k) -v_c(k)$, where $u_c(k)$ and $v_c(k)$ are velocity components of an ocean current. The velocity of the ocean current as well as the uncertainty associated with the measurement of the system states, $\bm{e}(k)$, were treated as disturbance signals. Note that quadratic damping was only considered in surge, where the vessel was assumed to reach the highest speeds. Except for this, the model is in agreement with the well-adopted framework found in~\cite{fossen2011handbook} and the definition of a second-order modulus model in~\cite{ljungberg2022a}. The parameter values, $\mathcal{X}_u =  -0.06$, $\mathcal{X}_{u|u|} = -0.01$, $\mathcal{X}_{vr} = 0.08$, $\mathcal{X}_{\tau} = 1.4 \cdot 10^{-5}$,  $\mathcal{Y}_v = -0.1$, $\mathcal{Y}_{ur} = -0.006$,  $\mathcal{Y}_{\tau} = 1.4 \cdot 10^{-5}$, $\mathcal{N}_r = -0.35$, $\mathcal{N}_{uv} = -0.03$, $\mathcal{N}_{\tau} = 3 \cdot 10^{-4}$, were chosen based on earlier work with the small-scale ship presented in Section~\ref{sec:experimental_study}, with a sampling rate of $f_s = 8$ Hz.

Data was generated based on 11 different excitation signals and the system states resulting from applying them in undisturbed simulations, $i.e.$, the expected experiment primitives, are given below.
\begin{enumerate}[label=(\subscript{\bm{\tau}}{{\arabic*}})]
	\item Decelerating motion: \\ $u(k) : 1 \to 0$, $v(k) = 0$, $r(k) = 0$.
	\item Accelerating motion: \\ $u(k) : 0 \to 1$, $v(k) = 0$, $r(k) = 0$.
	\item Slow/flat zig-zag motion \\ $u(k) \approx 0.35$, $v(k) \in [-0.1, 0.1]$, $r(k) \in [-0.2, 0.2]$.
	\item Moderate/flat zig-zag motion: \\ $u(k) \approx 0.8$, $v(k) \in [-0.1, 0.1]$, $r(k) \in [-0.2, 0.2]$.
	\item Fast/flat zig-zag motion: \\ $u(k) \approx 1.15$, $v(k) \in [-0.1, 0.1]$, $r(k) \in [-0.2, 0.2]$.
	\item Slow/steep zig-zag motion: \\ $u(k) \!\approx\! 0.35$, $v(k) \! \in \! [-0.15, 0.15]$, $r(k) \! \in \! [-0.35, 0.35]$.
	\item Moderate/steep zig-zag motion: \\ $u(k) \!\approx\! 0.7$, $v(k) \in [-0.15, 0.15]$, $r(k) \in [-0.35, 0.35]$.
	\item Fast/steep zig-zag motion: \\ $u(k) \approx 1$, $v(k) \in [-0.15, 0.15]$, $r(k) \in [-0.35, 0.35]$.
	\item Slow (inward-bound) spiral motion: \\ $u(k) \approx 0.4$, $v(k) \approx 0.05$, $r(k) : 0 \to -0.6$.
	\item Moderate (inward-bound) spiral motion: \\ $u(k) \approx 0.75$, $v(k) \approx 0.125$, $r(k) : 0 \to -0.6$.
	\item Fast (inward-bound) spiral motion: \\ $u(k) \approx 1$, $v(k) \approx 0.2$, $r(k) : 0 \to -0.6$.
\end{enumerate}
Here, all values for $u(k)$ and $v(k)$ are given in m/s and the values for $r(k)$ in rad/s. Each of the input signals were applied for about 40 s (300 samples) and the matrices in \eqref{eq:xbar}-\eqref{eq:zbar} were estimated. The problem \eqref{eq:opt_prob_zero_mean} was solved  in MATLAB using the function \textit{fmincon} and the interior-point algorithm, which resulted in an optimized design where $\bm{\tau_1}(k)$ is used for $16~\%$ of the total experiment time, $\bm{\tau_2}(k)$ for $4~\%$, $\bm{\tau_5}(k)$ for $9~\%$, $\bm{\tau_6}(k)$ for $42~\%$, $\bm{\tau_8}(k)$ for $4~\%$ and $\bm{\tau_{11}}(k)$ for $25~\%$. It can be noted that the most used experiment primitive in the optimized design is the slow and steep zig-zag motion associated with $\bm{\tau_6}(k)$. This is a bit surprising, considering that the signal magnitudes are strictly higher when input signals $\bm{\tau_7}(k)$ or $\bm{\tau_8}(k)$ are applied.

\begin{figure}
	\centering
	\includegraphics[width = 8.4cm]{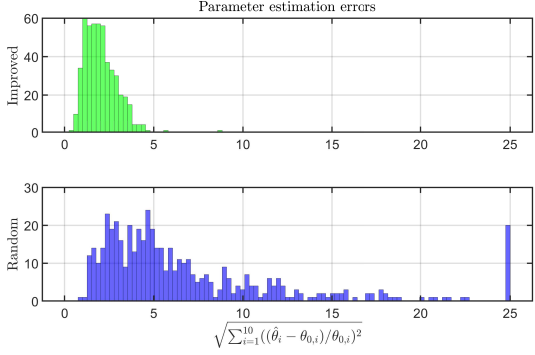}
	\caption{Norm of RMSE for normalized estimation errors between estimated parameters, $\hat{\bm{\theta}}$, and true parameters, $\bm{\theta}_0 = [\mathcal{X}_u \hdots  \mathcal{N}_{\tau}]^T$, over 500 Monte Carlo simulations. The top plot shows an optimized design and the bottom plot a random design. Errors above 25 are truncated to 25.}
	\label{fig:sim1}
\end{figure}

The predictor model
\begin{align} \label{eq:model_structure}
	\hat{\bm{y}}(k) = \begin{bmatrix} \bm{\varphi}^T_u(k) & 0 & 0 \\ 0 & \bm{\varphi}^T_v(k) & 0 \\ 0 & 0 & \bm{\varphi}^T_r(k) \end{bmatrix} \bm{\theta},
\end{align}
with
\begin{align*}
	\bm{\varphi}_u(k) &= \begin{bmatrix} y_1(k) & y_1(k) \abs{ y_1(k)} &  y_2(k)  y_3(k) & \tau_1(k) \end{bmatrix}^T,\\ 
	\bm{\varphi}_v(k) &= \begin{bmatrix} y_2(k)  &  y_1(k)  y_3(k) & \tau_2(k) \end{bmatrix}^T, \\ 
	\bm{\varphi}_r(k) &= \begin{bmatrix} y_3(k)  &  y_1(k)  y_2(k) & \tau_3(k) \end{bmatrix}^T,
\end{align*}
was used, which is able to describe the data-generating system in undisturbed conditions. As discussed earlier, the instruments were generated by simulating the output of this model with crude parameter values $\bm{\theta}'$ ($\theta'_1 = \theta'_5 = -0.08$, $\theta'_4 = \theta_7 = 10^{-5}$,  $\theta'_8 = -0.4$, $\theta'_{10} = 3.5 \cdot 10^{-4}$ and $\theta'_2 = \theta'_3 = \theta'_6 = \theta'_9 = 0$). As discussed before, the accuracy of an IV~estimator is highly dependent on the instruments and a too imprecise nominal model can make the estimator diverge. Despite neglecting all nonlinear effects, the nominal model above was sufficiently accurate to yield a reliable estimator. Further, an optimized input sequence of 125 s (1000 samples) was formed, which respected the found ratios and new data was collected where $u_c(k) \sim \mathcal{N}(0,0.025)$, $v_c(k) \sim \mathcal{N}(0,0.025)$ and $e_i(k) \sim \mathcal{N}(0,0.025)$ for $i = 1,2,3$. The models obtained from 500 Monte Carlo simulations using this optimized design were compared with models obtained using a random design. The random design was formed by arbitrarily choosing 1 out of the 11 input signals for each of 5 equally long segments (200 samples each). Which input signals that were used and in what order, was varied over the Monte Carlo simulations. For each Monte Carlo iteration, parameter estimation errors were computed and ideally, histograms showing these individually would be provided. Instead, due to limited space, a normalized root-mean-squared error (RMSE) for each estimated parameter was computed and the norm of all these is shown in Figure~\ref{fig:sim1}. By visual inspection it can be seen that the random design often gives higher estimation errors and that the improved design is a more robust choice in this regard. The improved design gives an error that is smaller than 5 in more than 99~\% of the cases, whereas the random design only does so in 49~\% of the cases.  

\begin{figure}
	\centering
	\includegraphics[width = 8.4cm]{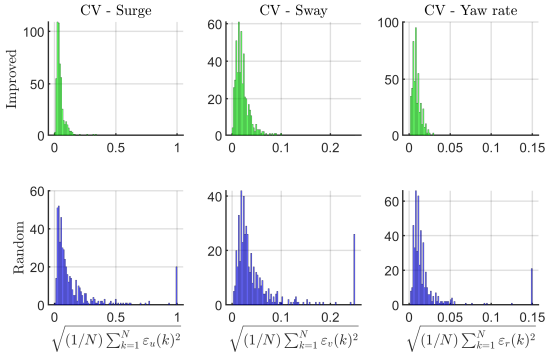}
	\caption{RMSE between simulated model output and output of the data-generating system for each DOF, $\varepsilon_u(k) = \hat{u}(k) - u(k)$, $\varepsilon_v(k) = \hat{v}(k) - v(k)$, $\varepsilon_r(k) = \hat{r}(k) - r(k)$. Errors above 1, 0.25 and 0.15 are truncated, respectively.}
	\label{fig:sim2}
\end{figure}

\begin{figure}
	\centering
	\includegraphics[width = 8.4cm]{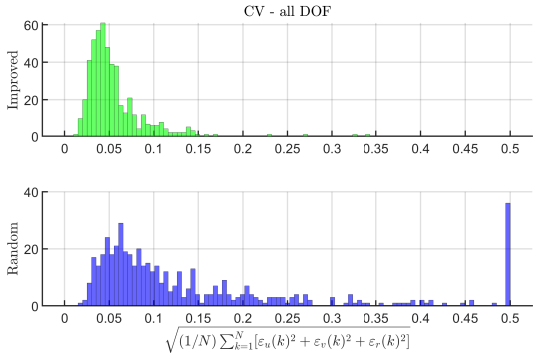}
	\caption{Norm of RMSE between simulated model output and output of the data-generating system, $\varepsilon_u(k) = \hat{u}(k) - u(k)$, $\varepsilon_v(k) = \hat{v}(k) - v(k)$, $\varepsilon_r(k) = \hat{r}(k) - r(k)$.  Errors above 0.5 are truncated to 0.5.}
	\label{fig:sim3}
\end{figure}

The accuracies of the estimated models were also evaluated by cross-validation (CV) between simulated model output and the output of the true system. In the validation dataset, an input signal that was different from all the experiment primitives was used. The RMSE metric was used in this case as well and in Figure~\ref{fig:sim2}, the errors for each DOF is shown and in Figure~\ref{fig:sim3}, the norm of these is shown. It can be seen that the simulated model output is more often in agreement with the system output when the optimized experiment design is used. Similar to when parameter errors were considered, it can in Figure~\ref{fig:sim3} be noted that the norm of the simulation errors is smaller than 0.15 in 99~\% of the cases when the optimized design is used but only in 69~\% of the cases when the random design is used. Similar observations can be made for each individual DOF in Figure~\ref{fig:sim2}.

\subsection{Results from model ship} \label{sec:experimental_study}
Experimental data was also collected using a small-scale surface ship, which is about 1 meter long, 0.3 meter wide and weighs roughly 14 kg. This ship is actuated by two azimuth thrusters mounted in the aft and a tunnel thruster located in the bow. Further, the ship is equipped with an inertial measurement unit (IMU) for orientation, a GNSS receiver for positioning, as well as rotational-speed sensors for the thrusters. The surge and sway speeds were estimated using Euler's explicit method with the position and orientation measurements. In order to make the analysis simple, the tunnel thruster was not used during the experiments and the input forces and moments, $\tau_1(k)$, $\tau_2(k)$ and $\tau_3(k)$, were generated solely based on the two azimuth thrusters. A simple static model for describing forces and moments obtained from azimuth thrusters is
\begin{align*}
	\tau_1(k) &= \sum_{i=1}^2 n_i(k) \cos\left(\alpha_i(k)\right), \\
	\tau_2(k) &= \sum_{i=1}^2 n_i(k) \sin\left(\alpha_i(k)\right), \\
	\tau_3(k) &= \sum_{i=1}^2 l_{y,i} n_i(k) \cos\left(\alpha_i(k)\right) + l_{x,i} n_i(k) \sin\left(\alpha_i(k)\right), \\
\end{align*}
where $l_{x,i}$ and $l_{y,i}$ define the (known) mounting position of azimuth thruster $i$ in $x$ and $y$-direction, respectively. The control signals, $n_1(k)$, $n_2(k)$, $\alpha_1(k)$ and $\alpha_2(k)$, were based on this model chosen to yield experiment primitives in agreement with the simulation experiments shown above. Moreover, the model structure was the same as in the simulation experiments and the instruments were  again generated by use of a nominal model. This time, the crude parameter values used for the nominal model were obtained by the least-squares estimate, $i.e.$, by letting $\bm{Z}(k) = \bm{\Phi}(k)$.

The collected data was divided into 55 parts (75 samples each), 5 for each of the 11 experiment primitives. Using these shorter datasets for estimating \eqref{eq:xbar}-\eqref{eq:zbar} and solving \eqref{eq:opt_prob_zero_mean}, gave an optimized design where $\bm{\tau_1}(k)$ is used for $20~\%$ of the total experiment time, $\bm{\tau_6}(k)$ for $49~\%$, $\bm{\tau_9}(k)$ for $13~\%$ and $\bm{\tau_{11}}(k)$ for $18~\%$. This optimized design was then approximated with a design where $\bm{\tau_6}(k)$ is used for half of the total time and $\bm{\tau_1}(k)$, $\bm{\tau_9}(k)$ and $\bm{\tau_{11}}(k)$ are used for one sixth of the time each. In this way, multiple different datasets could be formed from the collected data where the optimized ratios were fulfilled. The fact that multiple different realizations of the optimal experiment could be formed made it easier to discuss the results and to make statistical conclusions, something that is otherwise only possible in simulation experiments where more data is available.

\begin{figure}
	\centering
	\includegraphics[width = 8.4cm]{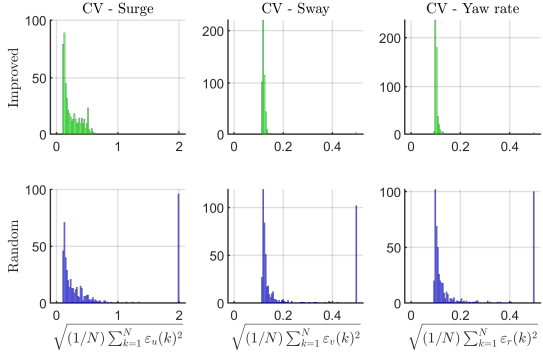}
	\caption{RMSE between simulated model output and estimates obtained from measured signals in each DOF from the small-scale model ship. Errors above 2, 0.5 and 0.5 are truncated, respectively.}
	\label{fig:model_boat1}
\end{figure}

\begin{figure}
	\centering
	\includegraphics[width = 8.4cm]{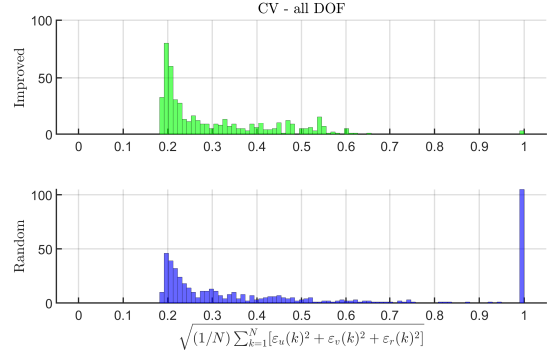}
	\caption{Norm of RMSE between simulated model output and estimates obtained from measured signals from the small-scale model ship. Errors above 1 are truncated to 1.}
	\label{fig:model_boat2}
\end{figure}

The optimized datasets were formed by picking 3 sub-experiments where $\bm{\tau_6}(k)$ is used and 1 sub-experiment where $\bm{\tau_1}(k)$, $\bm{\tau_9}(k)$ and $\bm{\tau_{11}}(k)$ are used each, $i.e.$, 6 shorter sub-experiments were used in total. This way of forming the experiment was compared to a random design, where 6 sub-experiments were selected randomly. Similarly to the simulation experiments, the accuracies of the estimated models were evaluated by CV between simulated model output and estimates obtained from the measured signals. For both the optimized design and the random design, the set of sub-experiments was varied 500 times and the results are given in Figures~\ref{fig:model_boat1} and \ref{fig:model_boat2}. It can be seen that the simulated model output is more often in agreement with the measured system output when the optimized experiment design is used. Here, it is most notable because the error following the random design takes on degenerate values in about 20~\% of the cases.

The validation data was made up by all the 55 sub-experiments, which means that the validation data was not completely independent of the estimation data ($6/55 \approx 11~\%$ overlap). This small overlap was deemed justifiable because the alternative would be a folded type of CV, where the sub-experiments currently used for estimation were kept out and not included in the validation set. Since there were only 5 sub-experiments of each type available, this would mean that some maneuvers occasionally were not present at all in the validation data and would therefore possibly result in a misleading evaluation.  

\subsection{Results from full-scale ship}
There was also data available from a full-scale ship. This ship is roughly 30 meters long and has an actuator setup with two azimuth thrusters mounted along the centerline, one at the front and one in the rear. This configurations makes it possible to excite the ship in sway without having a forward speed. The full-scale ship is equipped with a GNSS receiver with two antennas, which provides estimates of both its position and yaw angle. Unlike the small model ship discussed before, all signals were sampled at the lower rate of $f_s = 1$ Hz. Notably, the data from the full-scale ship was not collected for the sake of this work in particular and consequently the data does not include the previously discussed experiment primitives. Instead, the following motion types were identified.

\begin{enumerate}[label=(\subscript{\bm{\tau}}{{\arabic*}})]
	\setcounter{enumi}{11}
	\item Surge accelerations/decelerations: \\ $u(k) \in [0, 3.5]$, $v(k) = 0$, $r(k) = 0$.
	\item Sway accelerations/decelerations: \\  $u(k) = 0$, $v(k) \in [-1, 1]$, $r(k) = 0$.
	\item Slow sway motion: \\ $u(k) = 0$, $\abs{v(k)} \in [0.5, 0.8]$, $r(k) = 0$.
	\item Moderate sway motion: \\ $u(k) = 0$, $\abs{v(k)} \approx 0.9$, $r(k) = 0$.
	\item Fast sway motion: \\ $u(k) = 0$, $\abs{v(k)} \in [1.1, 1.2]$, $r(k) = 0$.
	\item Slow/high frequent ($T \approx 30$) zig-zag motion: \\ $u(k) \approx 3$, $v(k) \in [-0.5, 0.5]$, $r(k) = \in [-0.03, 0.03]$.
	\item Fast/high frequent ($T \approx 30$) zig-zag motion: \\ $u(k) \approx 4$, $v(k) \in [-0.5, 0.5]$, $r(k) = \in [-0.03, 0.03]$.
	\item Slow/low frequent ($T \approx 100$) zig-zag motion: \\ $u(k) \approx 3$, $v(k) \in [-0.5, 0.5]$, $r(k) = \in [-0.03, 0.03]$.
	\item Fast/low frequent ($T \approx 100$) zig-zag motion: \\ $u(k) \approx 4$, $v(k) \in [-0.5, 0.5]$, $r(k) = \in [-0.03, 0.03]$.
\end{enumerate}
As before, all values for $u(k)$ and $v(k)$ are given in m/s and the values for $r(k)$ in rad/s. Moreover, $T$ is the period time of the zig-zag motion in seconds. The simple model structure \eqref{eq:model_structure} turned out to be sufficient for describing the collected data and was therefore used here as well and the instrument matrix was generated in the same way as for the small-scale ship. Further, the data was divided into different parts for the different maneuvers. This time, 4 sub-experiments of each type were formed, each corresponding to 100 seconds of maneuvering ($i.e.$, 100 samples). Estimating \eqref{eq:xbar}-\eqref{eq:zbar} and solving \eqref{eq:opt_prob_zero_mean} gave an optimized design where $\bm{\tau_{13}}(k)$ is used for $4~\%$ of the total experiment time,  $\bm{\tau_{14}}(k)$ for $2~\%$, $\bm{\tau_{16}}(k)$ for $30~\%$, $\bm{\tau_{18}}(k)$ for $22~\%$, $\bm{\tau_{19}}(k)$ for $11~\%$ and $\bm{\tau_{20}}(k)$ for $31~\%$. 

\begin{figure}
	\centering
	\includegraphics[width = 8.4cm]{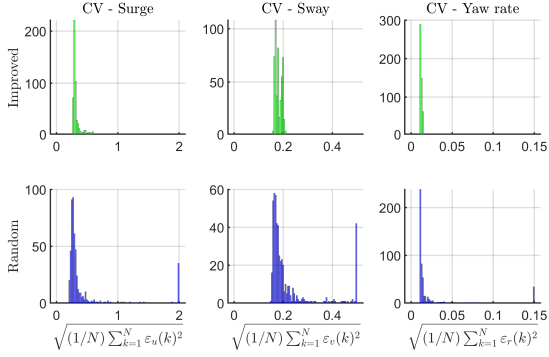}
	\caption{RMSE between simulated model output and estimates obtained from measured signals in each DOF from the full-scale ship. Errors above 2, 0.5 and 0.15 are truncated, respectively.}
	\label{fig:ship1}
\end{figure}

\begin{figure}
	\centering
	\includegraphics[width = 8.4cm]{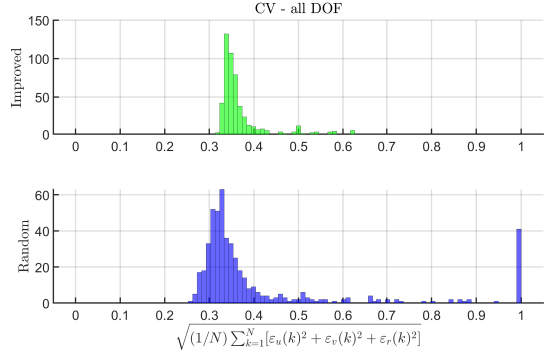}
	\caption{Norm of RMSE between simulated model output and estimates obtained from measured signals from the full-scale ship. Errors above 1 are truncated to 1.}
	\label{fig:ship2}
\end{figure}

Analogously to before, the optimal design was approximated with a design where $\bm{\tau_{16}}(k)$ and $\bm{\tau_{20}}(k)$ are used for one third each while $\bm{\tau_{18}}(k)$ and $\bm{\tau_{19}}(k)$ are used for one sixth each. In this case, realizations of an optimal experiment can be formed by picking 2 of the sub-experiments where $\bm{\tau_{16}}(k)$ is used, 1 where $\bm{\tau_{18}}(k)$ is used, 1 where $\bm{\tau_{19}}(k)$ is used and 2 where $\bm{\tau_{20}}(k)$ is used. Furthermore, the optimized experiment design was compared with a random design, where 6 sub-experiments were selected randomly and the accuracies of the estimated models were evaluated by CV between simulated model output and estimates obtained from the measured signals. For both the optimized design and the random design, the set of sub-experiments was varied 500 times and the results are given in Figures~\ref{fig:ship1} and \ref{fig:ship2}. Here it is actually the case that the most accurate models are obtained with the random design. The worst-case errors following the random design are, however, also higher and sometimes take on degenerate values. In this sense, the optimized design is a more solid choice.

It can be remarked that the data from the full-scale ship was collected on two separate occasions with quite different weather conditions. Parts of the data was collected during a day with wind speeds of less than 3 m/s, $i.e.$ light breeze by the definition in \cite{fossen2011handbook}, whereas the other parts were collected during a day where the wind speed was about 10 m/s, which corresponds to fresh breeze by the same set of definitions. Further, $\bm{\tau_{16}}(k)$-$\bm{\tau_{18}}(k)$ were only applied during the windy day and consequently, the data associated with those experiment primitives is more noisy in comparison to the other data. Due to the way that the realizations of the optimal design are formed, by concatenation of sub-experiments where both  $\bm{\tau_{16}}(k)$ and $\bm{\tau_{18}}(k)$ are used, they are bound to always include at least $50~\%$ of data from the windy day. The realizations of the random design are, however, occasionally made up exclusively of data from the calmer day, which partly explains why the most accurate models are obtained with the random design.

\section{Generating an experiment trajectory} \label{sec:gen_traj}
Most previous work about experiment design have focused on obtaining an optimal input signal. There are, however, many practical considerations to be made for a data-collection experiment. In addition to being time-efficient, it is, for example, important that the experiment is planned with respect to geographical constraints and operational safety. The feasibility of the resulting experiment trajectory is therefore equally important as the informativity content of the collected data. Motion planning is a vast field of research and the goal of this work was not to develop new theory in that regard. Instead, the focus is on showing how the previously discussed dictionary-based method for finding an excitation signal can be combined with a motion-planning framework to obtain a trajectory that is both informative and feasible.

In motion planning, a state-lattice-based approach uses a library of motion primitives that are combined to obtain a path from a starting position to a goal position, see for example \cite{Pivtoraiko2009}. Similarly to how the candidate excitation signals, $\bm{u}_1(k), ..., \bm{u}_Q(k)$, previously have been associated with information gains, each candidate signal can by use of a nominal model, or by initial experiments with the real platform, be associated with an expected change in position and attitude. These motion segments can then be used in lattice-based motion planning to find a suitable realization under constraints, such as spatial limitations or safety. The standard formulation of a lattice-based motion-planning problem is
\begin{subequations} \label{eq:kristoffer_motion_planning}
	\begin{align}
		& \underset{\{\bm{m}_k\}_{k=0}^{M-1}, M}{\text{minmize}}
		& & \sum_{k=0}^{M-1} J(\bm{m}_k) \\
		& \text{ s.t.}
		& & \bm{x}_0 = \bm{x}_s, \ \ \bm{x}_M = \bm{x}_f, \label{eq:constr1} \\
		& & & \bm{x}_{k+1} = f(\bm{x}_k, \bm{m}_k),  \label{eq:constr2} \\
		& & & \bm{m}_k \in \mathcal{P}, \label{eq:constr3} \\
		& & & c(\bm{m}_k, \bm{x}_k) \in \bm{\mathcal{X}}_{\text{free}}. \label{eq:constr4}
	\end{align}
\end{subequations}
adopted from \cite{bergman2021exploiting}. Here, $J(\bm{m}_k)$ is a cost function and and the decision variables are the motion primitive sequence $\{\bm{m}_k \}_{k=1}^M$ and its length $M$. Further, the constraints in \eqref{eq:constr1} define the initial and final states, the state-transition in \eqref{eq:constr2} gives the successor state $\bm{x}_{k+1}$ after $\bm{m}_k$ is applied in $\bm{x}_k$, the set $\mathcal{P}$ in \eqref{eq:constr3} includes all applicable motion primitives and lastly the constraint \eqref{eq:constr4} ensures that there is no collision with obstacles when $\bm{m}_k$ is applied in $\bm{x}_k$. 

Normally, the state vector contains information about position and attitude, which for motion planning in the horizontal plane includes positional coordinates, $(x,y)$, and heading angle, $\psi$. Therefore, introduce the notation $\bm{\eta}_k  = [x_k, \ y_k, \ \psi_k]^T$. The key difference between motion planning for dictionary-based experiment design and regular motion planning lies in the requirement of respecting the ratios between $N_1, \hdots, N_Q$ found when solving \eqref{eq:opt_prob} or \eqref{eq:opt_prob_zero_mean}. There are multiple ways to include such a constraint in the problem \eqref{eq:kristoffer_motion_planning}. One solution is to augment the state vector with additional elements that count how many times each experiment primitive has been used. This approach is conceptually appealing because of its interpretability. Thereby, let $g^q_k$ be the number of times experiment primitive $q$ has been used at time $k$. Including this, the state vector becomes
\begin{align}
	\bm{x}_k \overset{\Delta}{=} \begin{bmatrix} \bm{\eta}_k^T & g^1_k & \hdots & g^Q_k \end{bmatrix}^T.
\end{align}
The ratios between $N_1, \hdots, N_Q$ can then be respected by enforcing that each experiment primitive is used a chosen number of times
\begin{align}
	\bm{x}_s &= \begin{bmatrix} \bm{\eta}_s \\ 0 \\ \vdots \\ 0 \end{bmatrix}, \ \ \ \bm{x}_f = \begin{bmatrix} \bm{\eta}_G \\ n_1 \\ \vdots \\ n_Q \end{bmatrix}.
\end{align} 
The number of times experiment primitive $q$ is applied in total can be selected such that $n_q N'_q \approx N_q$. In this case, $N_1', \hdots, N_Q'$ become design variables, which declare how long each motion primitive should be. Due to the augmented state vector, the state lattice will be of higher dimension in comparison to a conventional lattice-based motion-planning problem and if too many experiment primitives (large $Q$) are considered and each is divided into a large number of parts (small  $N_1', \hdots, N_Q'$), the state lattice will grow quickly. Consequently, the trade-off between feasibility and computational complexity that follows the choice of $N_1', \hdots, N_Q'$ is important. Moreover, there are some general difficulties associated with a dictionary-based experiment design. This is because different concatenation orders will lead to different signal transitions, which affect the informativity of the design. Also, applying different signal types after each other might lead to abrupt changes and consequently high-frequency transients. These problems become more apparent when the experiment primitives are divided into smaller parts. To summarize, large values of $N_1', \hdots, N_Q'$ make the motion-planning problem easier to solve whereas small values of $N_1', \hdots, N_Q'$ give better possibilities of finding feasible solutions but increase the computational complexity and make the expected information gain from the experiment harder to predict. From a more practical point of view, some sub-experiments, such as zig-zag maneuvers, have natural breaking points at the end of each period. Choosing $N_1', \hdots, N_Q'$ in agreement with these is therefore a reasonable choice.

The objective function 
\begin{align}
	J = \sum_{k=0}^{M-1} J(\bm{m}_k),
\end{align}
is made up of the accumulated cost of applying a sequence of $M$ motion primitives. In a conventional motion-planning problem, the cost of applying a particular motion primitive can, for example, be related to travel time or fuel cost. Performing motion planning for experiment design is a bit different. Ideally, a data-collection experiment should only be made up by information-yielding maneuvers, but in practice this is not always possible. This is because it is sometimes necessary to reposition the ship prior to executing the next maneuver due to space constraints. One way of dealing with this issue is to assume that there are additional basic motion primitives that can be used to connect the informative sequences. In this case, the complete set of motion primitives is 
\begin{align}
	\mathcal{P} = \left\{ \underbrace{\bm{m}^1, \hdots, \bm{m}^Q}_{\text{informative}}, \underbrace{\bm{m}^{Q+1}, \hdots, \bm{m}^{Q+B}}_{\text{basic}}\right\},
\end{align}
where the basic motion primitives can be of types that are often used in conventional motion planning, such as straight driving as well as plain left and right turns. The goal with the experiment design can then be formalized as using each information-yielding primitive a given number of times, while using the basic primitives as sparsely as possible. A simple objective function that caters to this goal is
\begin{align} \label{eq:objective_function_planning}
	J(\bm{m}^{\ell}) = \begin{cases} 0 \ &\text{if} \ \ell \leq Q, \\
		\bar{L} \ &\text{if} \ \ell > Q.
	\end{cases}
\end{align}
In other words, have the cost of applying an information-yielding primitive be 0 and the cost of applying a basic primitive be $\bar{L} > 0$. 

For high-dimensional and large state lattices, an exhaustive search for a solution is time consuming. In this case, the planning can be carried out using the $A^*$-search algorithm, as formulated in \cite{bergman2021exploiting}. The online computational efficiency of such a search method relies on a well-informed heuristic function to guide the search towards the solution. The heuristic function takes as input two arbitrary states in the state lattice and returns an estimated cost of going from one to the other. Typically, the second input argument is the goal node, such that the heuristic function returns the estimated cost of reaching the goal. In general, it is only possible to guarantee optimality if the heuristic function is both admissible and consistent. It is, however, not easy to come up with a non-trivial heuristic function that fulfills these criteria for the planning problem considered in this work. This is because the cost of reaching the goal node from another node can be 0, even if this other node happens to be far away from the goal in terms of geographical distance. In fact, considering \eqref{eq:objective_function_planning}, the cost will be 0 if the goal position can be reached by using any combination of the information-yielding experiment primitives that are still left to be used. This means that it is hard to say for sure that a solution is optimal unless most other options have been explored.

\subsection{Simulation example}
To show that the proposed way of generating an experiment trajectory is reasonable, a simulation experiment was performed based on a sea-depth map over the port of Helsinki. Three different experiment primitives were considered, two types of zig-zag motion in surge direction with low and high amplitude for the turns, respectively, as well as one zig-zag motion in sway direction. Each of these were applied 3 times. In addition to the information-yielding maneuvers, the ship was allowed to move straight forward and to rotate on the spot (both clockwise and anticlockwise), $i.e.$, 3 basic motion primitives were considered as well. The obstacles were represented by bounding boxes and additional bounding boxes associated with the changes in position from the motion primitives were computed. The constraint \eqref{eq:constr4} could then be checked by examining whether the bounding box associated with a particular motion primitive overlapped with any bounding box associated with an obstacle. The state lattice was discretized with $35 \times 44$ positional coordinates and 4 different attitude levels. In this simulation experiment, a heuristic function with three terms was implemented. The first two terms were based on the Euclidian distance and deviation in attitude with respect to the goal, which constitutes standard costs in motion planning. The third term was an artificial cost associated with the remaining required use of the information-yielding experiment primitives. By giving significant weight to the third term, a greedy behavior where the planner prioritized using information-yielding maneuvers early was obtained. This improved the speed of the search for a solution. The considered heuristic function was neither admissible nor consistent but gave good results in practice. 

\begin{figure}
	\centering
	\includegraphics[width = 8.4cm]{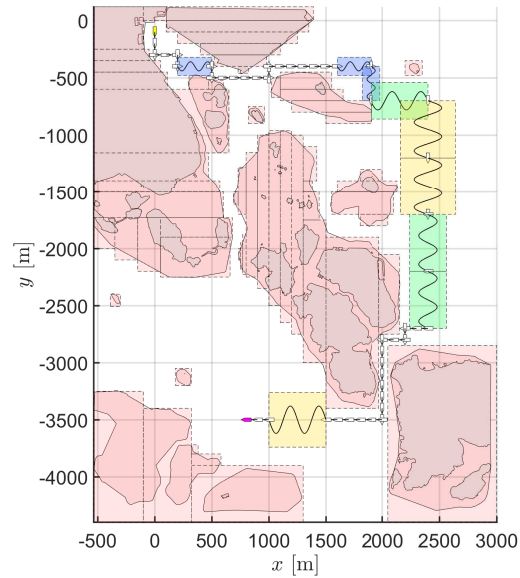}
	\caption{Planned trajectory using 3 different experiment primitives; steep zig-zag motion in surge direction (blue), wide zig-zag motion in surge direction (orange) and wide zig-zag motion in sway (green). The start position is marked in yellow and the goal position is marked in purple. Islands and shallow water are treated as obstacles and are marked in~red.}
	\label{fig:traj1}
\end{figure}

A planned trajectory is shown in Figure~\ref{fig:traj1}, where the ship was assumed to start in the upper-left corner, which is a harbor and to finish up in a position in the lower parts of the map. The goal position was chosen arbitrarily and an arguably more realistic scenario would be to both start and end in the harbor. Planning with the same node as start and goal is, however, by no means more difficult and choosing the goal node in this way had the benefit of the trajectory not overlapping itself, which makes the result much easier to illustrate in a plot. It can be seen that the planner chooses to use the most space-consuming maneuvers in the wide area at the top right. Further, the importance of including the basic motion primitives is clear, because without them it would, for example, not be possible to pass by the narrow region around (2000, -3500). Finding this trajectory took about 40~seconds on a modern laptop when using a completely plain implementation of the $A^*$-search algorithm in MATLAB. A more state-of-the-art implementation would probably speed up the search significantly. 

\section{Conclusions} \label{sec:conc}
The usefulness of a dictionary-based approach for experiment design has been analyzed and promising results have been shown on both simulated and real data from marine vessels. Further, a technique for combining the approach with a motion-planning framework to obtain an experiment trajectory that is both informative and feasible has been demonstrated. Selecting a set of informative maneuvers in the proposed way can probably in itself give improvements in terms of model accuracy in comparison to what is currently done in practice and if a more generally optimal input signal is sought, the suggested method can be viewed as a way of obtaining an initial guess for warm-starting a more flexible optimizer. Evaluating the potential benefits of such an approach is relevant future work.

\begin{ack}                               
This work was supported by the competence center LINK-SIC.  
\end{ack}

\bibliographystyle{plain}        
\bibliography{bib}           	 
%
%
\end{document}